\documentclass[12pt]{article}
\input epsf
\newcommand{\be}{\begin{equation}}
\newcommand{\ee}{\end{equation}}
\newcommand{\bea}{\begin{eqnarray}}
\newcommand{\eea}{\end{eqnarray}}
\newcommand{\ba}{\begin{array}}
\newcommand{\p}[1]{(\ref{#1})}
\newcommand{\ea}{\end{array}}

\def\bbox{{\,\lower0.9pt\vbox{\hrule \hbox{\vrule height 0.2 cm
\hskip 0.2 cm \vrule height 0.2 cm}\hrule}\,}}
\newcommand{\dsl}{\pa \kern-0.5em /}

\newcommand{\nn}{\nonumber \\}

\def\ds{\raise.15ex\hbox{/}\kern-.57em\partial}
\def\Ds{\,\raise.15ex\hbox{/}\mkern-13.5mu D}
\begin{document}

\baselineskip 18pt


\begin{titlepage}
\vfill
\begin{flushright}
DAMTP-2000-98\\
HUTP-00/A038\\
KIAS-P00057\\
hep-th/0009061\\
\end{flushright}

\vfill

\begin{center}
\baselineskip=16pt 
{\Large\bf String Fluid from Unstable D-branes} \\
\vskip 10.mm
{Gary Gibbons$^{+1}$, Kentaro Hori$^{*2}$, and Piljin Yi$^{\#3}$ } \\
\vskip 0.5cm
{\small\it
$^+$D.A.M.T.P., C.M.S., Cambridge University, Wilberforce Road,
\\ Cambridge CB3 0WA, U.K.}
\vskip 0.5cm
{\small\it
$^*$ Jefferson Physical Laboratory, Harvard University,
\\ Cambridge, MA 02138, USA}
\vskip 0.5cm
{\small\it
$^\#$School of Physics, Korea Institute for Advanced Study\\
207-43, Cheongryangri-Dong, Dongdaemun-Gu, Seoul 130-012, Korea}
\end{center}
\vfill
\par
\begin{center}
{\bf ABSTRACT}
\end{center}
\begin{quote}
We consider Sen's effective action for unstable D-branes, and study its 
classical dynamics exactly. In the true vacuum, the Hamiltonian dynamics 
remains well-defined despite a vanishing action, and is that of massive 
relativistic string fluid of freely moving electric flux lines. The 
energy(tension) density equals the flux density in the local co-moving 
frame. Furthermore, a finite dual Lagrangian exists and is related to 
the Nielsen-Olesen field  theory of ``dual'' strings, supplemented by 
a crucial constraint. We conclude with discussion on the endpoint of
tachyon condensation.
\end{quote}

\vfill
\vskip 5mm
\hrule width 5.cm
\vskip 5mm
\begin{quote}
{\small
\noindent $^1$ E-mail: G.W.Gibbons@damtp.cam.ac.uk\\
$^2$ E-mail: hori@infeld.harvard.edu\\
$^3$ E-mail: piljin@kias.re.kr\\
}
\end{quote}
\end{titlepage}
\setcounter{equation}{0}

\section{Introduction}

When an unstable D-brane or a pair of D-brane and anti-D-brane decays, 
all open string states should disappear. Around the new true vacuum 
of the theory where brane tension is all dissipated away, it follows, 
the theory must consist of closed string states only. Exactly how 
this is achieved from the open string point of view, however, 
is far from clear. In particular, very little is known about whether
any open string degrees of freedom survive, say, in the form of light
coherent or solitonic states, and if so how they are related to the 
closed string degrees of freedom. One possibility is that such states
correspond to usual closed string modes repopulating the worldvolume.
It is the purpose of this note to shed some light one some of these issues.

There has been incremental progress in understanding
how open string state might be removed from
the spectrum. The pioneering work by A. Sen addressed the fate of some 
massless open string modes \cite{sen}. 
In particular, in a D-brane and anti-D-brane
system, one linear combination of massless gauge fields was shown to 
acquire a massgap via an Abelian Higgs mechanism. While this finite mass 
seemed to fall short of removing the state altogether, it is a step toward 
the right direction. 

This naturally leads to  the other linear combination of gauge fields
that has no charged fields in perturbative open string sector, which 
posed extra difficulty in understanding how it was  removed 
from the spectrum \cite{sred,K}. 
On the other hand, while no perturbative state 
is charged under $U(1)$, there are nonperturbative
states that carry magnetic charge. This gives one a possibility that a dual 
Higgs mechanism exists, induces a massgap, and confines the $U(1)$ gauge 
field \cite{piljin}. 
For instance, one finds precisely this nonperturbative behavior from 
S-dualizing the perturbative sector of D3-anti-D3 system.

One drawback of the picture is that details of the scenario have been checked 
largely in the strong coupling picture. The practical matter of how one may 
study the mechanism in the weak coupling regime, beyond the formal 
manipulation such as introducing gauge-invariant mass term for dual gauge 
field, still remains. However, regardless of such details, it seems clear 
that the confined flux strings must emerge in any regime, if one is to believe
in the charge conservation; for the electric flux of this $U(1)$ carries
the fundamental string charges \cite{piljin2}.

Understanding of this gauge field at weak coupling were considerably
improved with the introduction of an effective action by Sen 
\cite{unstable,potential}.
The tachyon potential is identified with the vanishing tension of 
the decaying branes which appears in front of the Born-Infeld action, 
\be
-V(T)\sqrt{-Det\,(\eta+{\cal F})}, \label{sen}
\ee
and sets the inverse coupling of the gauge field in question. 
The subsequent large coupling would certainly confine charged
particles, if there were any. However, this by itself does not quite
explain how the gauge field might be removed. The charge
confinement here is more like that in the classical QED$_{2+1}$, 
where the confinement is simply due to the behavior of the Green's function. 
For instance, no massgap is to be found.\footnote{Recent 
computations \cite{wati,abaa} suggest that an explicit
mass term might be present in the tree-level open string field theory,
as opposed to (\ref{sen}). See the last section of \cite{wati}
for a discussion on this issue.
See also \cite{j} for a related computation.
For a more recent discussion see \cite{Asen}.}
It is here where the dual or magnetic objects
can play their most important role.
They can generate a mass gap, correctly
quantize the electric flux and confine it
into a thin tube.
In particular, it was shown \cite{piljin2}
in the $2+1$ dimensional case, where the dual ``objects'' generate
the potential term for the dual photon \`a la Polyakov,
that the electric flux tube tends to be completely
squeezed and the tension approaches the exact value of the fundamental
strings.

While this $U(1)$ gauge field is only one of infinitely many open string
modes, proper understanding of its fate may lead to something far more
interesting than how open string modes disappear. To see this, it suffices to
recall that ${\cal F}$ is {\it the} open string mode that carries the 
fundamental string charge, and its dynamics may as well play a role in
the true vacuum which is supposed to be a closed string theory.

In this note, we will take the classical Lagrangian of the general form
\p{sen}, and study the dynamics in some detail. First we reformulate 
the theory in the Hamiltonian formalism. We must emphasize that this is not
a sterile mathematical exercise, the crucial point being that, in the
limit of $V=0$, the correct physical degrees of freedoms appear naturally
in the Hamiltonian approach \cite{piljin2}. From general form of Hamiltonian, 
we learn that the theory reduced to that of stringy fluid of
electric flux lines. A finite and fully relativistic, dual formulation 
is shown to exist, and used to study the stringy behavior of the theory.
Finally we discuss on possible interpretations.

\section{Canonical Formulation of Born-Infeld Theory}

The Lagrangian is
\be
{\cal L}= -V(T)\sqrt{-Det(\eta+{\cal F}+\partial X^I\partial X_I)},
\ee
where $X^I$, $I=1,\dots, D-d$ are transverse scalars, with the spacetime 
dimension $D$, and 
${\cal F}_{\mu\nu}=\partial_\mu A_\nu-\partial_\nu A_\mu$ is the curvature
tensor of the gauge field. We will adopt the mostly positive signature 
for the metric. While we put the tachyonic potential $V(T)$ in front, 
the analysis in this section is applicable when $V$ is replaced by the 
constant tension. A more recent proposal \cite{garouberg} 
places
tachyon kinetic term inside the determinant on equal footing as $X^I$.
This possibility is also covered by the analysis in this section.

The determinant can be expressed more conveniently in terms of $E_i\equiv 
{\cal F}_{0i}$ and $F_{ij}={\cal F}_{ij}$ as
\be
-Det(\eta+F+\nabla X^I\nabla X_I)=Det(h)(1-\dot X^I\dot X_I)
-E^{(+)}_iD_{ik}E^{(-)}_k,
\ee
where we further defined 
\be
h_{ik}\equiv \delta_{ik}+F_{ik}+\partial_i  X^I\partial_k X_I ,
\ee
and the matrix $D$ as
\be
D_{ik}=(-1)^{i+k}\Delta_{ki}(h),
\ee
where $\Delta_{ki}$ denotes the determinant of the matrix with the $k^{th}$ 
row and the $i^{th}$ column omitted. When $h$ is invertible, this is equivalent
to
\be
D= Det(h) h^{-1}.
\ee
We also introduced a shorthand notation 
\be
E^{(\pm)}_i \equiv E_i \pm \dot X^I\partial_i X_I .
\ee

Now that we isolated the $E$-dependence of the Lagrangian, it is a
straightforward exercise to find the canonical variables and find the
Hamiltonian. Conjugate momenta are,
\be
\pi^i=\frac{\delta {\cal L}}{\delta \dot A_i}
=\frac{V}{\sqrt{-Det(\eta+{\cal F}+\partial X^I\partial X_I)}}
\frac{E^{(+)}_kD_{ki}+D_{ik}E^{(-)}_k}{2},
\ee
which satisfy the Gauss constraint $\partial_i\pi^i=0$, and
\be
p_I=\frac{\delta {\cal L}}{\delta \dot X^I}
=\frac{V}{\sqrt{-Det(\eta+{\cal F}+\partial X^I\partial X_I)}}
\left(Det(h)\dot X_I -\frac{E^{(+)}_kD_{ki}-
D_{ik}E^{(-)}_k}{2}\,\partial_iX_I \right),
\ee
which are conjugate to $X^I$. The Hamiltonian consistent with the Gauss
constraint is obtained by the following Legendre transformation 
\be
{\cal H}=\pi^iE_i+p_I\dot X^I-{\cal L},
\ee
which yields
\be
{\cal H}=
\frac{VDet(h)}{\sqrt{-Det(\eta+{\cal F}+\partial X^I\partial X_I)}}.
\ee
This Hamiltonian, when expressed in terms of the
canonical variables, becomes
\be
{\cal H}=\sqrt{ \pi^i\pi^i +p_Ip^I+ (\pi^i\partial_iX^I)^2 +(F_{ij}\pi^j+
\partial_iX^Ip_I)^2 +V^2Det(h)}
\ee
As mentioned at the top of the section, 
this computation is carried over straightforwardly when we include a
tachyon kinetic term inside the square-root, \`a la Bergshoeff et.al. All we
need to do is to introduce another scalar $T$ on equal footing as 
$X^I$. 

An important set of physical quantities are the conserved Noether momenta.
The time translation invariance leads to the conserved energy, ${\cal H}$,
while the spatial translation leads to conserved momentum,
\be
{\cal P}_{i}= -F_{ik}\pi^k-\partial_iX^Ip_I,
\ee
where we covariantized the Noether momentum by adding a total derivative.

As an aside, we note that the Hamiltonian can be written formally as
\be
\sqrt{\pi^M\pi^M+{\cal P}_M{\cal P}_M +V^2Det(h)}, \label{D}
\ee
where $\pi^M$ is conjugate to $(A_M)=(A_i, X_I)$, and the ``conserved 
momenta'' ${\cal P}_M$ is
\be
{\cal P}_{N}= -F_{NM}\pi^M,
\ee
with $F_{NM}=\partial_N A_M -\partial_M A_N$ with $\partial_I\equiv 0$.
The potential piece $Det(h)$ can also be written in a similar fashion
\cite{T};
\be
Det(\delta_{ij}+F_{ij}+\partial_iX^I\partial_jX^I)=Det(\delta_{NM}+
F_{NM}).
\ee
This reflects the underlying T-duality, and provides a
consistency check of the Hamiltonian we derived.

\section{Electric Flux after Tachyon Condensation}

Suppose we started with an unstable brane with some electric flux
on it, and let it decay via tachyon condensation. When the tachyon 
condenses to its true ground state where the potential $V(T)$ vanishes
\cite{potential,osft,emil}, 
the Lagrangian vanishes. Despite this rather violent process, the
electric flux must be preserved somehow. A crucial point we need to
note is that
the conserved electric flux is not given by $E_i$ but rather by $\pi^i$.
The situation is analogous to that of massless relativistic particle.
The Lagrangian $L=-m\sqrt{1-v^2}$ vanishes in the limit $m\to 0$
but the Hamiltonian $H=\sqrt{p^2+m^2}$ survives the limit.
(The analogy is more or less precise in the case of unstable D1-brane,
as shown in \cite{piljin2} and later repeated in \cite{chicago}.)
The necessity of use of $\pi^i$ rather than $E_i$
can also be seen in several ways. One way is to observe that  $\pi^i$ 
is the quantity that is constrained by the Gauss constraint. Perhaps a 
more intuitive way is to recall that $\pi^i$ is the quantity that acts
as the source for NS-NS spacetime two-form field $B$ associated with the 
fundamental strings. This is because $B$ appears in the Born-Infeld
action in the combination $F+\hat B$ where $\hat B$ denotes the pull-back
of $B$ to the worldvolume. In other words, $\pi^i$ carries
the spacetime fundamental string charge which has to be preserved no matter
what.

At the end of decay process, when the D-brane tension is 
dissipated by emission of closed string states, the Hamiltonian 
is obtained by setting $V=0$,
\be
{\cal H}=\sqrt{ \pi^i\pi^i +p_Ip^I+ (\pi^i\partial_iX^I)^2 
+{\cal P}_i{\cal P}_i }.
\ee
Given a total flux, the part of energy that cannot be dissipated away is
the one associated with $\pi$. When all other excitations vanish, the 
energy of the system is simply
\be
|\pi|=\sqrt{\pi^2}.
\ee
To minimize the energy, the flux line wants to be straight and directed
along one direction. Once this is achieved, however, the energetics does not
care how the flux lines are distributed. In the co-moving frame of straight and
parallel flux lines, the energy per 
unit length equals the total flux \cite{piljin2}.

For flux lines moving on the worldvolume, 
we can define a Lorentz-invariant quantity $\tau$ that 
reduces to $|\pi|$ in the local co-moving frame,
\be
\tau=\pi^2/{\cal H}.
\ee
The invariance property will be shown in section 5.
Consider the effect of turning on momenta ${\cal P}$. The energy is,
\be
{\cal H}=\sqrt{\pi^i\pi^i+{\cal P}_i{\cal P}_i }
=\sqrt{\pi^2+{\cal H}^2v^2 },
\ee
where $v_i\equiv {\cal P}_i/{\cal H}$ is the velocity field. 
The energy density of the moving flux lines is
\be
{\cal H}=\tau\times\frac{1}{1-v^2}.
\ee
The energy density of the moving flux lines are equal to boosted energy 
density. In other words, the flux lines behave as massive relativistic 
string fluid.

Turning on the transverse scalars as a perturbation, we find another
interesting aspect of the flux line dynamics. When $\pi$ is the dominant
quantity, we may expand $p_I$ to find
\be
p_I\simeq -{\cal H}\dot X_I.
\ee
When the flux lines move out of the worldvolume with a uniform 
velocity, then, the energy associated with a uniform motion becomes
\be
{\cal H}\simeq \tau\times \frac{1}{1-v^2-(\dot X^I)^2}.
\ee
This suggests that the transverse motion is also relativistic and  
on equal footing as motions within worldvolume.

When the flux string bends and  stretches along the transverse directions,
this turn on $(\pi^i\partial_iX^I)^2$ term, upon which
an additional term in the Hamiltonian,
\be
{\cal H}=\sqrt{\pi^2(1+(\partial_\parallel X^I)^2)+\cdots}.
\ee
This precisely takes into account of the fact that the flux line has been
lengthened. The effective tension density is again $|\pi|$ measured in 
the co-moving frame. Therefore, we conclude that the flux lines behave as if
they are a continuum of massive relativistic strings that moves in all
of spacetime.

\section{Dual Description and Equation of Motion}

We saw that the dynamics is perfectly finite 
despite the vanishing action. All this says, on the other hand,
is that the coordinate variables we started with, namely the $U(1)$ gauge
field, are not necessarily good variables.
A natural followup question is whether there exists
another set of variables which are good in this limit of $V=0$.
For the remainder of the note, we will ignore the transverse scalars.
Inclusion of them would proceed trivially along the same spirit as in
\p{D}, without extra work.

A standard trick \cite{arkady}
is to dualize the fields where one chooses to treat 
the magnetic fields, $F_{ij}$, to be the conjugate momenta of 
the new, dual variables. Performing Legendre transformation on $F_{ij}$,
one finds,
\be
{\cal L}'={\cal H}-\frac12 F_{ij}K^{ij}=\sqrt{\pi^2-K^2/2},
\ee
with
\be
K^{ij}=2\frac{\delta {\cal H}}{\delta F_{ij}}=
\frac{1}{\cal H}\,\left(F_{im}\pi^m\pi_j-\pi_iF_{jm}\pi_m\right). \label{K}
\ee
The underlying, dual variables are $(d-3)$-form potential $C$, whose 
$(d-2)$-form field strength $G=dC$ is related to $\pi$ and $K$ via,
\be
{\cal K} \equiv \pi_i\,dt\wedge dx^i+ \frac12 \,K_{ij}dx^i\wedge dx^j=
*G,\label{hodge}
\ee
where $*$ is the Hodge-dual operator. The Legendre transformation gives
\be
{\cal L}'=\sqrt{-{\cal K}^2/2}=\sqrt{G^2/2}.
\ee
Note that $(G_{0\cdots})^2$ comes with the negative sign, so that stable
field configurations are those with dominant magnetic $G$-fields, or
equivalently those with dominant electric ${\cal K}$-field.

However, this ${\cal L}'$ is not quite what we want. A subtlety arises 
from the fact that the Hamiltonian $\cal H$ has flat directions in $F_{ij}$;
$K^{ij}$ as a function of $F_{ij}$ is rather restricted, and must be 
parallel to $\pi_i$. In particular ${\cal K}$ obeys
the constraint ${\cal K}\wedge{\cal K}=0$.
The subtlety then manifests itself in a pathology 
that the canonical analysis of ${\cal L}'$ does not lead us back to 
the Hamiltonian $\cal H$. For this, there is a simple cure;
we only need to impose a constraint via a Lagrange multiplier\footnote{
All fields here must have an origin in string theory, yet 
the origin of $\mu$ appears a bit mysterious; No worldvolume 
field of such a type is apparent in the perturbative open string sector. 
One massless tensor that would be of rank $(d-4)$ is the ``would-be
Goldstone boson'' associated with the dual magnetic objects. In the
present limit, it could indeed appear as a nondynamical field.
Whether and why it should have an axion-like coupling is unclear, however.} 
$(d-4)$-form $\mu$ that imposes the constraint
${\cal K}\wedge {\cal K}=0$;
\be
{\cal L}_{\rm dual}=\sqrt{G^2/2} +\langle *\mu, *G\wedge *G\rangle/4.
\label{dual}
\ee
The equation of motion is
\be
0=d \left(\frac{{\cal K}}{\sqrt{-{\cal K}^2/2}}+*(\mu\wedge{\cal K})\right),
\label{eom}
\ee
which, in terms of the field strength ${\cal F}_{\mu\nu}$
of the original variable, is the Bianchi identity $d{\cal F}=0$.
The Bianchi identity for $G$, $dG=0$, which reads as
\be
\partial^{\mu}{\cal K}_{\mu\nu}=0
\label{b}
\ee
is equivalent to the equation of motion and the Gauss constraint of the
original description. The derivation of these statements
is rather technical and therefore is recorded in Appendix.

The constraint thus imposed $*G\wedge *G={\cal K} \wedge {\cal K}=0$ has a 
rather simple physical interpretation; it tells us that, at each point, 
the two-form ${\cal K}$ is proportional to  an area element of a 
two-dimensional surface. Furthermore, the stability requirement that the
Lagrangian is real, tells us that the surfaces must be timelike 
(${\cal K}^2<0$). From
the definition of ${\cal K}$ above, the family of these ``surfaces'' is
clearly the foliation of the worldvolume by the flux fluid.

The equation of motion (\ref{eom}) admits a scaling symmetry
\bea
&&{\cal K}\rightarrow f{\cal K},\nn
&&\mu \rightarrow \mu /f,
\label{scale}
\eea
for any reasonable function $f$. 
On the other hand, the Bianchi identity (\ref{b}) for $G$
restricts admissible form of $\lambda$ to be those satisfying
\be
0=(\partial_\mu f){\cal K}^{\mu\nu}.
\ee
In other words, $f$ is allowed to vary only in the plane orthogonal 
to the flux fluid. For any static solution with electric flux $\pi_i$ only,
this generates other solutions of the form $f\pi_i$ as long as the
function satisfies
\be
\pi^i\partial_i f=0 .
\ee
Nothing in the classical dynamics favors a particular distribution of
flux lines along any orthogonal direction.

The above form of the action, {\it without} the constraint, was first 
discussed by Nielsen and Olesen \cite{nielsen}, in 4-dimensional setting. 
They observed that the equation of motion admits 
magnetic $G$-flux strings as special solutions of the form,
\be 
{\cal K}^{\mu\nu}(x)= \int \delta^{(d)}(x-Z(\sigma))\,dZ^\mu\wedge dZ^\nu,
\ee
where $Z(\sigma)$ is a map from the worldsheet coordinates $\sigma^{0,1}$
to the worldvolume, and the integral is over $\sigma^{0,1}$.
These solutions satisfy the constraint by construction, so they are also 
solutions of \p{eom} with $\mu=0$. Nielsen and Olesen also showed
that these solutions behave exactly like a Nambu-Goto
string. However, it is quite clear that this confined form of the flux 
lines is  an artifact of the ansatz, as the symmetry \p{scale} shows.

\section{Energy-Momentum and Fluid Motion}

This dual action allows an easy computation of the energy momentum tensor, 
which comes out to be quite simple;
\be
T_{\mu\nu}
=2\frac{\delta}{\delta g^{\mu\nu}}\left(\sqrt{g}\sqrt{g^{\alpha_1\beta_1}
\cdots g^{\alpha_{d-2}\beta_{d-2}} G_{\alpha_1\cdots\alpha_{d-2}}
G_{\beta_1\cdots\beta_{d-2}}}\right)
=\frac{{\cal K}_{\mu\lambda}{\cal K}_\nu^{\;\;\lambda}}{\sqrt{-{\cal K}^2/2}}.
\label{emte}
\ee
The contribution from the Lagrange multiplier goes away once the constraint
is used. It is straightfoward to check the conservation of the energy-momentum,
$\partial^\mu T_{\mu\nu}=0$, using \p{eom}, \p{b}, and the constraint.

The temporal part $T_{0\mu}$ reproduces the energy density and 
the momentum density faithfully,
\bea
&&T_{00}=\frac{\pi^2}{\sqrt{\pi^2-K^2/2}}=\sqrt{\pi^2+{\cal P}^2}={\cal H},\nn
&&T_{0i}=\frac{\pi^m K_{im}}{\sqrt{\pi^2-K^2/2}} 
= F_{ij}\pi^j =-{\cal P}_i,
\eea
while the stress-tensor is equally simple;
\be
T_{ij}=(-\pi_i\pi_j
+{\cal P}_i{\cal P}_j)/{\cal H}=\frac{-\pi_i\pi_j
+{\cal P}_i{\cal P}_j}{\sqrt{\pi^2+{\cal P}^2}}.
\ee
It is fairly easy to see how this form of energy-momentum would
arise from a fluid of flux lines. Suppose we started with a  bundle
of straight flux lines, say pointing toward direction 1.  Let the distribution
in the co-moving frame be characterized by a Lorentz scalar function, 
which we call $\tau'$, so that the rest frame  energy-momentum is
\be
\left(\begin{array}{cccc}\tau' &   0  & 0 & \cdots \\
                       0 & -\tau' & 0 & \cdots \\
                       0 &  0    & 0 & \cdots \\
                  \vdots & \vdots &\vdots &\vdots
\end{array}\right).
\ee    
Suppose we boosted the system along a direction orthogonal to the bundle, 
say direction 2,
with speed $v$. The Lorentz transformation takes this energy-momentum tensor 
to
\be
\left(\begin{array}{cccc}\tau'\gamma^2 &   0  & -\tau'\gamma^2v & \cdots \\
                       0 & -\tau' & 0 & \cdots \\
                      -\tau'\gamma^2v&  0    & \tau'\gamma^2v^2 & \cdots \\
                  \vdots & \vdots &\vdots &\vdots
\end{array}\right),
\ee    
with the dilation factor $\gamma=1/\sqrt{1-v^2}$. Identifying the energy 
density $\tau'\gamma^2$ with $\cal H$ and the momentum density 
$\tau'\gamma^2v$
with $\cal P$, we see that the local form of the above energy-momentum 
tensor is reproduced precisely. It is a matter of identifying $\pi_i/|\pi|$
with the direction 1, and ${\cal P}_i/|{\cal P}|$ (which is necessarily 
orthogonal to $\pi_i$) with direction 2. As a final consistency check, one 
finds that $\tau'$ is correctly mapped to a Lorentz invariant quantity;
\be
\tau'={\cal H}/\gamma^2
={\cal H}\left(1-({\cal P}/{\cal H})^2\right)=\frac{\pi^2}{\cal H}=\tau,
\ee
which we promised earlier to be Lorentz invariant. Indeed,
\be 
\tau=-\frac12\,{T_\mu^{\;\;\mu}}=\sqrt{-{\cal K}^2/2},
\ee
is a Lorentz scalar. In the co-moving frame, therefore, the only force 
is the pull along the flux lines coming from this tension density
$\tau$. The theory reduced to that of  flux lines with no mutual
pressure.

For the sake of completeness, we record here the equations of
motion for the flux lines. The conservation of the energy-momentum 
combined with the $G$-Bianchi identity produces,
\bea
\partial_0 n^i +v^k\partial_k n^i=n^k\partial_k v^i, &&\nn
\partial_0 v^i +v^k\partial_k v^i=n^k\partial_k n^i, &&\label{eof}
\eea
where the vector fields $n$ and $v$ are defined as
\bea
\pi={\cal H}n, &&\nn
{\cal P}={\cal H}v,
\eea
which satisfies the constraints
\bea
n^in^i+v^iv^i=1,\qquad n^iv^i=0.
\eea
The evolution of the energy density ${\cal H}$ is then determined via,
\be
\partial_0{\cal H}+v^k\partial_k{\cal H}+{\cal H}(\partial_kv^k)=0,
\ee
supplemented by an initial condition satisfying $0=\partial_i\pi^i
=\partial_i\left({\cal H}n^i\right)$. The equations of motion
\p{eof} have exceedingly simple physical interpretations. The first is 
merely a continuity equation; the flux lines bend in response to the 
gradient in the velocity field. The second, dynamical equation says 
that the velocity fields adjust itself to straighten flux lines. 
The latter again confirms that in the comoving frame, the only 
acceleration arises from the tension.

The system of a 2-form field
${\cal K}$ constrained by ${\cal K}\wedge{\cal K}=0$
and ${\cal K}_{[\rho\lambda}\partial^{\mu}{\cal K}_{\nu]\mu}=0$ 
with the energy momentum tensor (\ref{emte}) has been considered in
\cite{sta,let} and called string dust or string cloud model.
(The second constraint follows from the $G$-Bianchi identity (\ref{b}).)

\section{Discussion}

We analyzed Sen's effective action for unstable D-branes, and found
that, classically, the gauge system reduces exactly to a theory of flux
fluid which carries fundamental string charge density. Tension density,
electric flux density, and fundamental string charge density all coincide.
The theory contains Nambu-Goto strings as special solutions, but does 
not naturally form confined strings of unit flux.
The symmetry \p{scale} 
of the equation of motion pretty much guarantees this. 

This brings us back to the original question of what is the endpoint of the
tachyon condensation. In full  string theories, there is little doubt
that the true vacuum is some sort of closed string theory. For 
non-spacetime-filling branes, the unstable brane system can be thought
of as a classical lump of energy {\it in a closed string theory}, 
and all that happen is the lump dissipates away to infinity, 
leaving behind a vacuum of the same closed string theory. 

In order to obtain Sen's effective action, however, one must perform
several truncations of the theory. First one considers the open string
sector rather than the closed string theory. Here the original unstable
branes cannot be thought of as classical lump anymore. One then truncates to
the massless fields plus the tachyon, and also takes the classical limit of 
the theory. Note that the effective coupling of the truncated theory is 
not the string coupling but the string coupling divided by $V$. This 
somewhat violent limit must be responsible for the 
appearance of the string fluid.

The interpretation of this flux fluid remains to be found. The most
cautious approach would be to view the classical theory as valid only 
in the macroscopic settings. If one cannot distinguish distribution of
fundamental strings from continuum electric flux, and he may identify 
one with the other. Whatever the microscopic mechanism of forming the
fundamental strings, macroscopic observers find the fundamental strings 
that moves in the entire spacetime, rather than just inside the D-brane 
worldvolume. This would be the correct interpretation if the classical 
dynamics is unable to resolve a length scale arbitrarily larger than 
$\sqrt{\alpha'}$, and if the unit flux of a fundamental
string is viewed as infinitesimal.

A better interpretation would be that we are allowed, within the classical 
dynamics, to distinguish between isolated unit flux tube and well-dispersed
distribution. If this is the case, we have a natural framework to
address the question of how fundamental string is recovered from
the originally dispersed flux lines. As emphasized above, the classical
system does not care how flux lines are distributed in a rest 
frame.\footnote{It was recently suggested \cite{Asen} in different contexts 
that such a degeneracy might be a result of inappropriate choice of
coordinate variables. 
However, the infinite degeneracy here occurs in the profile of {\it energy 
density}. Energy density does make sense in the effective field theory we 
are considering and should not depend on the choice of coordinates. }
While the smallest of perturbation in the gauge sector
could render the flux lines repulsive or attractive,
such a purely energetical mechanism is unlikely to produce strings
of precise unit flux. What one needs is {\it quantization of flux in
local sense, rather than in global sense}, which points toward a 
mechanism involving a dual object.

A dual (i.e., magnetic) Higgs mechanism could do the trick and 
confine the flux into strings of
unit quantum quite naturally \cite{piljin}. In the language 
of section 4, this would generate a local, gauge-invariant, 
Higgs-like mass term of the dual field $C$ \cite{piljin,rey}, and
force the flux lines to 
come together. A candidate magnetic object has been identified for,
say, the D-brane and anti-D-brane system:
open D-brane of two less dimensions
ending on the pair has the right charge.
Due to the nature of the classical
string fluid (that {\it any} profile
in the transverse direction has the same total energy),
a slight effect of the dual object can squeeze
the electric flux into a very thin tube.
As in the $2+1$ dimensional case analyzed in \cite{piljin2},
if that happens the tension of the flux string
approaches the exact value of the fundamental string tension.
(In \cite{piljin2}, it was claimed for $p\geq 3$ cases
that it was necessary to assume that the scale associated with the dual
Higgs mechanism is of order $1$ in the string coupling.
However, it is now quite clear that one does not need such an
assumption, given the lack of classical 
interactions between adjacent flux lines.)

Alternatively, one may think of the dual object
as local deformation of worldvolumes of the D-branes, accompanied by flux 
lines. When extended, the magnetic object is heavy and 
unsuitable for weak-coupling description, but they 
could reappear as light worldvolume fields once the decay process commences 
and fluctuations of D-brane become cheap.

An interesting  generalization of this strong-coupling limit
of Born-Infeld theory is for the M-theory branes.
An unstable configuration of an M5-anti-M5 is expected to decay and produce
co-dimension-three worldvolume solitons to be identified with M2-brane.
It is not unreasonable to expect a membrane fluid dynamics in the limit
where the five-brane tension is dissipated away. The worldvolume
theory need not be chiral in this case, and we should find a simple 
dual description analgous to \p{dual}. This study may provide us more 
information on M-theory branes.

Finally we would like to comment that
there have been several studies of Born-Infeld action in the limit of
vanishing tension or (effectively) infinite $\alpha'$. Ref.~\cite{UBI}
studies the latter limit for D3, in effect, from which we benefited
much. There were more recent studies of tensionless D-branes in 
Ref.~\cite{lindstrom}, which also found string-like
degrees of freedom. It is unclear to us at the moment how the latter
work is related to ours.

\medskip

\section*{Acknowledgement}

We would like to thank P. Kraus, A. Lawrence, J. Maldacena, E. Martinec,
S. Minwalla, S. Shatashvili, and S. Rey for helpful conversations.
We are grateful to Aspen Center for Physics (K.H., P.Y.) and to CERN
Theory Group (G.G., P.Y.) for warm hospitality.
The research of K.H. is supported in part by NSF-DMS 9709694.

\section*{Appendix: Equations of Motion and Bianchi Identities}
\makeatletter
\renewcommand{\theequation}{A.\arabic{equation}}
\@addtoreset{equation}{section}
\makeatother
\setcounter{equation}{0}

We record here some of the details in the algebraic manipulations
used in section 4.
In particular, we show here that the $G$-Bianchi identity is the
$\cal F$-equation of motion (including the Gauss constraint), and the
$G$-equation of motion is the $\cal F$-Bianchi identity.
As in section 4, we set all $X^I=0$ and the Lagrangian is
\be
{\cal L}=-V\sqrt{Det(h)-E^T DE}.
\ee
The conjugate momenta simplify to
\be
\pi=\frac{V(D+D^T)E}{2\sqrt{Det(h)-E^T DE}},
\ee
and obey the Gauss law constraint $\partial_i\pi_i=0$.
The Hamiltonian is given by
\be
{\cal H}=\sqrt{\pi^i\pi^i+(F_{ij}\pi^j)^2+V^2Det(h)}=
{VDet(h)\over \sqrt{Det(h)-E^TDE}}.
\ee
Useful formulae to note are
\bea
&&(1-F)\pi={VDE\over \sqrt{Det(h)-E^T DE}},
\nonumber\\
&&(1+F)\pi={VD^TE\over \sqrt{Det(h)-E^T DE}},
\nonumber\\
&&(1-F^2)\pi={VDet(h)E\over \sqrt{Det(h)-E^TDE}}.
\nn
\eea
Using these formulae it is straightforward to show that the variation of
the Lagrangian is
\be
\delta{\cal L}
=\delta E^T\pi-{V^2\over 2{\cal H}}{\rm Tr}(D\delta F)
+{1\over {\cal H}}\pi^T\delta FF\pi.
\label{delL}
\ee
We drop the second term in the $V\to 0$ limit where the Hamiltonian is
given by
\be
{\cal H}=\sqrt{\pi^2+(F\pi)^2}.
\ee
Then, the Euler-Lagrange equations read as
(the first one is the Gauss constraint)
\bea
&&\partial_i\pi_i=0,\\
&&\partial_0\pi_i+\partial_j\left[{1\over {\cal H}}
\Bigl(\pi_j(F\pi)_i-\pi_i(F\pi)_j\Bigr)\right]=0.
\label{Feom}
\eea
The Bianchi identity for ${\cal F}$ is expressed in terms of $\pi^i$ and
$F_{ij}$ as
\bea
&& \partial_{[i}F_{jk]}=0,\nn
&& \partial_0 F_{ij} = 
\partial_i\left(\frac{\pi_j - F_{jk}F_{kl}\pi_l}{\cal H}\right)
-\partial_j\left(\frac{\pi_i - F_{ik}F_{kl}\pi_l}{\cal H}\right).
\label{Fb}
\eea
where we traded off $E_i$ in favor of the physical field $\pi_i$.

Let us see whether the Bianchi identity and equation of motion for the
dual side reproduces these equations. Let us consider the equation of
motion first. First let us define
\be
{\cal J}_{\mu\nu}:={{\cal K}_{\mu\nu}\over \sqrt{-{\cal K}^2/2}}+
\Bigl[*(\mu\wedge{\cal K})\Bigr]_{\mu\nu}.
\ee
If $\mu$ is written in terms of a 4-form $\alpha$ as $\mu=*\alpha$
and if we denote $J_{ij}={\cal J}_{ij}$,
this decomposes into
\bea
&&{\cal J}_{0i}={{\cal K}_{0i}\over \sqrt{-{\cal K}^2/2}}
-{1\over 2}\alpha_{0ijk}{\cal K}_{jk},
\label{Jdef1}\\
&&J_{ij}={{\cal K}_{ij}\over \sqrt{-{\cal K}^2/2}}
+\alpha_{0ijk}{\cal K}_{0k}
-{1\over 2}\alpha_{ijkl}{\cal K}_{kl}.
\label{Jdef2}
\eea
{}From the constraint ${\cal K}\wedge{\cal K}=0$ and the bound
$-{\cal K}^2/2\geq 0$, one can put
\bea
&&{\cal K}_{0i}=a_i,
\\
&&{\cal K}_{ij}=-a_ib_j+a_jb_i,
\eea
with $a\cdot b=0$ and $-{\cal K}^2/2=a^2(1-b^2)\geq 0$. 
Taking the contractions of (\ref{Jdef2}) with $a_i$ and $b_i$ and using
(\ref{Jdef1}),
we obtain
\bea
&&b_i={J_{ij}a_j\over \sqrt{a^2+(Ja)^2}},
\\
&&{\cal J}_{0i}={a_i-(J^2a)_i\over\sqrt{a^2+(Ja)^2}}.
\eea
Then, the equation of motion (\ref{eom}) reads as
\bea
&&\partial_{[i}J_{jk]}=0,
\\
&&\partial_0J_{ij}=
\partial_i\left({a_j-(J^2a)_j\over\sqrt{a^2+(Ja)^2}}\right)
-\partial_j\left({a_i-(J^2a)_i\over\sqrt{a^2+(Ja)^2}}\right),
\eea
while the Bianchi identity (\ref{b}) amounts to
\bea
&&\partial_i a_i=0,
\\
&&\partial_0 a_i
+\partial_j\left[{a_j(Ja)_i-a_i(Ja)_j\over\sqrt{a^2+(Ja)^2}}\right]=0.
\eea
We see that these equations are equivalent to the ${\cal F}$-Bianchi
identity (\ref{Fb}) and ${\cal F}$-equation of motion (\ref{Feom})
under the identification
\be
a_i=\pi_i,\qquad J_{ij}=F_{ij}.
\ee
It is easy to see that this identification is consistent with 
the dualization of variables, \p{K} and \p{hodge}. 

This identification of the two descriptions determines the classical
value of the Lagrange multiplier to be of the form,
\be
\mu=*\left(\frac{dt\wedge \pi_idx^i\wedge F}{\pi^2}+ 
dt\wedge \lambda +\nu\right),
\ee
where 4-form $\nu$ is a purely spatial and otherwise arbitrary, 
while $\lambda$ is appropriately constrained by 4-form $\nu$,
up to a 3-form orthogonal to $\pi$. Since the Lagrange multiplier is
not part of the physical phase space, this ambiguity is harmless
and may be ignored.


\vskip 1cm

\begin{thebibliography}{99}



\bibitem{sen}
A.~Sen,
``Tachyon condensation on the brane anti-brane system'',
JHEP {\bf 08} (1998) 012, hep-th/9805170;

\bibitem{sred}
M.~Srednicki,
``IIB or not IIB'',
JHEP {\bf 9808} (1998) 005, hep-th/9807138.

\bibitem{K}
E.~Witten, 
``D-Branes and K-Theory'', 
JHEP {\bf 9812} (1998) 019,
hep-th/9810188.

\bibitem{piljin}
P.~Yi,
``Membranes from five-branes and fundamental strings from Dp branes'',
Nucl.\ Phys.\ {\bf B550} (1999) 214,
hep-th/9901159.

\bibitem{piljin2}
O.~Bergman, K.~Hori, and P.~Yi,
``Confinement on the Brane'',
Nucl.\ Phys.\ {\bf B580} (2000) 289,
hep-th/0002223.

\bibitem{unstable} A.~Sen,
``BPS D-branes on non-supersymmetric cycles'',
JHEP {\bf 9812}, 021 (1998)
[hep-th/9812031];
``Non-BPS states and branes in string theory'',
hep-th/9904207.

\bibitem{potential}
A.~Sen,
``Supersymmetric world-volume action for non-BPS D-brane''",
JHEP {\bf 9910} (1999) 008, hep-th/9909062;
"Universality of the tachyon potential", 
JHEP {\bf 9912} (1999) 027, hep-th/9911116.




\bibitem{wati}
W. Taylor, ``Mass generation from tachyon condensation for 
vector fields on D-branes'', hep-th/0008033.

\bibitem{abaa}
A.~Sen and B.~Zwiebach,
``Large marginal deformations in string field theory'',
hep-th/0007153;
A.~Iqbal and A.~Naqvi,
``On marginal deformations in superstring field theory'',
hep-th/0008127.

\bibitem{j}
J.~R.~David,
``U(1) gauge invariance from open string field theory'',
hep-th/0005085.

\bibitem{Asen}
A.~Sen,
``Some Issues in Non-commutative Tachyon Condensation'',
hep-th/0009038.

\bibitem{garouberg}
M.~R.~Garousi,
``Tachyon couplings on non-BPS D-branes and Dirac-Born-Infeld action'',
Nucl.\ Phys.\  {\bf B584} (2000) 284
[hep-th/0003122];
E.A. Bergshoeff, M. de Roo, T.C. de Wit, E. Eyras, S. Panda,
``T-duality and Actions for Non-BPS D-branes'',
JHEP {\bf 0005} (2000) 009, hep-th/0003221.

\bibitem{T}
G. W. Gibbons, ``Born-Infeld particles and Dirichlet p-branes'',
Nucl. Phys. {\bf B514} (1998) 603.

\bibitem{osft} A.~Sen and B.~Zwiebach,
``Tachyon condensation in string field theory'',
JHEP {\bf 0003} (2000) 002, hep-th/9912249; N.~Berkovits, 
``The tachyon potential in open Neveu-Schwarz string field theory'',
hep-th/0001084; N.~Berkovits, A.~Sen, and B.~Zwiebach, 
``Tachyon condensation in superstring field theory'',
hep-th/0002211; N. Moeller and W. Taylor, ``Level truncation 
and the tachyon in open bosonic string field theory''
Nucl. Phys. {\bf B583} (2000) 105-144, hep-th/0002237;
P.~De Smet and J.~Raeymaekers,
``Level four approximation to the tachyon potential
in superstring field  theory'',
JHEP {\bf 0005} (2000) 051,
hep-th/0003220;
A.~Iqbal and A.~Naqvi,
``Tachyon condensation on a non-BPS D-brane'',
hep-th/0004015.
L.~Rastelli and B.~Zwiebach,
``Tachyon potentials, star products and universality'',
hep-th/0006240.

\bibitem{emil}     
J. A. Harvey, D. Kutasov, E. J. Martinec,
``On the relevance of tachyons'', hep-th/0003101.


\bibitem{chicago}
J. A. Harvey, P. Kraus, F. Larsen, E. J. Martinec
``D-branes and Strings as Non-commutative Solitons'',
JHEP {\bf 0007} (2000) 042, hep-th/0005031.

\bibitem{arkady}
A.~A.~Tseytlin, 
``Self-duality of Born-Infeld action and Dirichlet 
3-brane of type IIB superstring theory'',
Nucl.\ Phys.\  {\bf B469}, 51 (1996), hep-th/9602064.

\bibitem{nielsen}
H.B. Nielsen and P. Olesen, ``Local field theory of the dual string'',
Nucl. Phys. {\bf B57} (1973) 367.

\bibitem{sta}
J.~Stachel,
``Thickening The String. I. The String Perfect Dust'',
Phys.\ Rev.\  {\bf D21} (1980) 2171.

\bibitem{let}
P.~S.~Letelier,
``Clouds Of Strings In General Relativity'',
Phys. Rev. {\bf D20} (1979) 1294;


\bibitem{rey}
S. Rey, 
``The Higgs Mechanism for Kalb-Ramond Gauge Field'',
Phys. Rev. {\bf D40} (1989) 3396.

\bibitem{UBI}I. Bia\l ynicki-Birula, ``Nonlinear-electrodynamics: 
Variations on a theme by Born and Infeld,'' in {\it Quantum Theory 
of Fields and Particles}, Eds.B Jancewicz and J Lukierski, World 
Scientific, Singapore (1983); ``Field theory of photon dust,'' 
Acta Physica Polonica {\bf B 23} (1992) 553.


\bibitem{lindstrom}
U. Lindstrom and R. von Unge, ``A Picture of D-branes at 
Strong Coupling'', Phys. Lett. {\bf B403} (1997) 233,
hep-th/9704051;
H. Gustafsson and U. Lindstrom, ``A Picture of D-branes at Strong Coupling II. 
Spinning Partons, Phys. Lett. {\bf B440} (1998) 43,
hep-th/9807064;
U. Lindstrom, M. Zabzine, and  A. Zheltukhin,
"Limits of the D-brane action'', JHEP {\bf 9912} (1999) 016,
hep-th/9910159 






\end{thebibliography}
\end{document}